\documentclass[fleqn,10pt]{wlscirep}
\usepackage[utf8]{inputenc}
\usepackage[T1]{fontenc}
\usepackage{verbatim}
\usepackage{pdflscape}
\usepackage{afterpage}
\usepackage{soul}
\usepackage[hypcap=false]{caption}

\title{Sustainable Development Goals as unifying narratives in large UK firms' Twitter discussions}
\date{A beautiful spring in 2022}

\author[1,2,3*]{Alessia Patuelli}
\author[4,5,2]{Fabio Saracco}

\affil[1]{Northumbria University, Amsterdam Campus, Corry Tendeloo Building (CTH), Fraijlemaborg 133, 1102CV Amsterdam, The Netherlands}
\affil[2]{IMT School for Advanced Studies Lucca, p.zza S. Francesco 19, 55100, Lucca, Italy}
\affil[3]{Department of Economics and Management, University of Florence, Florence, 50127, Italy}
\affil[4]{Institute for Applied Computing, National Research Council of Italy,  via dei Taurini 19, 00185 Rome, Italy}
\affil[5]{`Enrico Fermi' Research Center (CREF), via Panisperna 89A, 00184 Rome, Italy}

\affil[*]{alessia.patuelli@northumbria.ac.uk}

\newcommand\fab[1]{{\color{black}#1}}
\newcommand\fabX[1]{{\color{black}#1}}
\newcommand\aleX[1]{{\color{black}#1}}
\newcommand\aleY[1]{{\color{black}#1}}
\newcommand\ale[1]{{\color{black}#1}}

\begin{abstract}
\aleX{
To achieve sustainable development worldwide, the United Nations set 17 Sustainable Development Goals (SDGs) for humanity to reach by 2030. Society is involved in the challenge, with firms playing a crucial role. Thus, a key question is to what extent firms \aleY{engage with}  the SDGs. Efforts to map firms’ contributions have mainly focused on analysing companies’ reports based on limited samples and non-real-time data. We present a novel interdisciplinary approach based on analysing big data from an online social network (Twitter) with complex network methods from statistical physics. By doing so, we provide a comprehensive and nearly real-time picture of firms’ engagement with SDGs. 
Results show that: 1) SDGs themes tie conversations among major UK firms together; 2) the social dimension is predominant; 3) the attention to different SDGs themes varies depending on the community and sector firms belong to; 4) stakeholder engagement is higher on posts related to global challenges compared to general ones; 5) large UK companies and stakeholders generally behave differently from Italian ones. 
This paper provides theoretical contributions and practical implications relevant to firms, policymakers and management education. Most importantly, it provides a novel tool and a set of keywords to monitor the influence of the private sector on the implementation of the 2030 Agenda.

}
\end{abstract}
\begin{document}

\flushbottom
\maketitle
%
%
\thispagestyle{empty}


\section*{Introduction}


\aleX{Reaching sustainable development is an urgent need for humanity. The term was conceptualised back in 1987 as “the development that meets the needs of the present without compromising the ability of future generations to meet their own needs” \cite{WCED1987}. Subsequent contributions from the United Nations (UN) continued to clarify it and to outline the dimensions of sustainable development \cite{Tsalis2020}. The latest effort was made in 2015. On 27 September 2015, the UN established 17 Sustainable Development Goals (SDGs) and 169 targets to be reached by a joint effort from all members of society by 2030. The goals balance three dimensions of sustainable development (economic, social, and environmental) and encourage action in areas vital for humanity and the world. Firms are considered crucial development players in achieving the SDGs \cite{scheyvens2016private}, while the goals are coherent with the concept of Corporate Social Responsibility (CSR). In fact, the three dimensions into which the SDGs can be grouped are coherent with the three dimensions of CSR \cite{elkington1997cannibals}:} the social dimension (1–5, 10, 16, and 17), the economic (7–9, 11, and 12), and the environmental one  (6, 13–15) \cite{d2021assessing}.
\aleX{The concept of CSR goes back to the 1950s and has been variously defined. A generally accepted definition refers to a company’s relationships and responsibilities to society, regarded as the groups of stakeholders with which it interacts \cite{dahlsrud2008corporate,carroll1999corporate,Snider2003}. }
It comprises all firms’ activities beyond what is required by law \cite{mcwilliams2001corporate}. What CSR means in practice varies on the cultural and historical environment in which a company operates, and may also represent the difficulties that a company is dealing with at the time \cite{Snider2003}.
Despite being primarily a societal phenomenon, SDGs have the potential to significantly advance CSR research \cite{bebbington2018achieving}, with CSR serving as a theoretical framework to examine how and to what degree businesses contribute to the SDGs \cite{Vildåsen2018256}. 
Stakeholder engagement is a related concept which has been differently defined and may be viewed under many different theoretical viewpoints. It has been conceptualised as “practices the organisation undertakes to involve stakeholders positively in organisational activities” \cite{greenwood2007stakeholder} (p. 315). \\
\aleX{Contributing to SDGs is a new challenge for companies worldwide \cite{Rosati2019}, which can significantly contribute to sustainable development \cite {VanZantenvanTulder2021}. One primary question is to what extent companies engage with global challenges, namely the SDGs. Papers have been developed with this aim, both conceptual and empirical. Please find a thorough discussion of the primary studies in the next section. The problem of capturing \fabX{all} firms’ contributions to SDGs is still unsolved, as papers mostly focus on small samples that cannot give a whole picture of the phenomenon. Plus, most papers base their analysis on companies’ reports, thus providing a non-real-time picture of companies’ contributions. 
Our study contributes to tackling this issue. }In this paper, we aim to answer the following broad research question: \textit{ To what extent are businesses engaging with SDGs themes?}. \\ 
\aleX{Understanding how businesses contribute to the SDGs is crucial for several reasons. First, we are now mid-way between when the SDGs were set (2015) and when they are aimed (2030). Enough time has passed to evaluate since the SDGs’ establishment, while there is still room for improvement in future times. 
\aleY{Second, businesses are crucial actors and, due to the urgent need to achieve sustainability worldwide \cite{UN2015}, it is essential to capture their engagement with global challenges, as defined by the SDGs \cite{VanZantenvanTulder2021}.}
Third, scholars believe that research in this area is still embryonic \cite{EMMA2021126781}. Fourth, it is essential to develop novel methods to describe} firms’ advancements towards SDGs with big data, a quick and low-cost tool \cite{Mio20203220,deVilliers2021598}. \\
Indeed, online social networks provide a new, underexploited tool to understand firms’ challenges, CSR activities and stakeholder engagement \cite{Patuelli2021}. In fact, in the last 15-20 years, online social networks have changed communication, making it cheaper and faster than before and providing a new channel for businesses to engage and directly interact with their stakeholders. They now represent a crucial means of disseminating firms’ CSR activities and involving stakeholders. Online social networks also provide the tools to measure stakeholder engagement, assuming that the users belong to the firm’s stakeholders \cite{bonson2013set}. Our research will also investigate stakeholder engagement with SDGs themes. 

Legitimacy theory and stakeholder theory are the two primary approaches \cite{Brown1998, Guthrie1989} that explain why companies are active in online social networks. On the one hand, legitimacy theory claims that businesses act following society's expectations and ideals. These are not constant and change across time and space. Although several scholars have related legitimacy theory to CSR, it is not necessarily restricted by CSR or stakeholder expectations. According to this perspective, firms use online social networks to justify their social position \cite{Manetti2016}. On the other hand, following stakeholder theory, firms should follow stakeholders' expectations to create long-term value. Consistently with this approach, firms utilise online social networks to communicate with their stakeholders and share their strategies and outcomes \cite{giacomini2020environmental}.
We base on Twitter, which is extensively used in online societal debates since its \fabX{short messages} are particularly suitable for \fabX{fast communication, as breaking news or political slogans. In fact, it has been used extensively for investigating political debates in different countries and how they are affected by disinformation (see 
\aleY{Ref.s~\cite{GonzalezBailon2013,Cresci2015, DelVicario2016c,Ferrara2017,Zollo2017,Shao2018a,Stella2018,Stella2019, Becatti2019c,Bovet2019,Gallotti2020,Cinelli2020,Caldarelli2020a,Caldarelli2021, Mattei2021,Guarino2021,Cinelli2021} for an incomplete, but almost exhaustive, review).}
The availability of detailed data permits a fine characterization of accounts and their engagement in discussions (see, for instance, Ref.s~\cite{GonzalezBailon2013,Stella2019}) and to highlight non-trivial structures and dynamics~\cite{DelVicario2016c,Zollo2017,Caldarelli2020a,Caldarelli2021, Mattei2021}}.
It is an excellent source for investigation, together with its data availability\fabX{, especially in countries, as the UK, where its adoption is particularly high (see the statistics in the following).} In contrast to interviews and survey-based data, this method does not rely on response rates or an individual’s desire to respond to get a bigger sample size. On Twitter, users (stakeholders) can retweet and like posts, which can be considered an endorsement of the message’s substance \cite{Conover2011, Saxton2019}.
\aleX{For these reasons, we believe that online social networks, specifically Twitter, are suitable for this study.}
We focus on large firms, as they have a high social impact \cite{Campopiano2015,Iaia2019} and are eager to engage in social and environmental activities \cite{giacomini2020environmental}, \aleX{although other factors, as the firms’ sector \cite{CALABRESE2022132324} and age \cite{Noorhayati2018}, have an impact.} Compared to small and medium enterprises, they often have more stakeholders requiring information \cite{Campopiano2015}. Thus, we focus on all large firms in one European country, the UK.
\aleX{As will be discussed in the following section, research on firms’ accomplishments towards SDGs has been based chiefly on companies’ reports. We propose a different approach, combining novel data sources for the management discipline and an interdisciplinary approach for the analysis. Our research is based on 5,859 accounts of large UK firms and 3.1M tweets posted between 2021/02/17 and 2022/02/17. These data are then analysed with complex network methods from statistical physics, showing the communities of discussion that naturally arise. \aleY{As further discussed in the Concluding Remarks, one limitation of this approach is that it only focuses on the communication dimension. It is beyond the scope of this paper to check to what extent companies are tackling the themes they are discussing on Twitter.}
In addition to our primary research question,} we compare our findings on the UK firms with analogous research investigating large Italian firms’ Twitter discussion and CSR orientation. Developing from the social and institutional paradigm, we expect that firms belonging to countries with different institutions, cultures, and values show different behaviours \cite{hofstede1984culture,GRAY1988}. 

\aleX{This article develops as follows. First, we provide a  review  on previous findings about firms' contributions to SDGs, providing an overview of the main contributions in the field. Then, we proceed with the results, describing firms' engagement with SDGs topics, the communities of discussions that arise and the engagement of stakeholders on these issues. We continue presenting some concluding remarks, contributions, practical implications and future research paths. Last, we detail our method. }

\section*{Previous results}
\aleX{SDGs were established in September 2015 as global objectives for our societies. Instead of focusing on a macro-perspective \cite{Rotondo2022}, this paper investigates the micro-dimensions, i.e. firms’ contributions to global challenges. The UN explicitly indicates firms as key players in working towards the SDGs: “Private business activity, investment and innovation are major drivers of productivity, inclusive economic growth and job creation. We acknowledge the diversity of the private sector, ranging from micro-enterprises to cooperatives to multinationals. We call upon all businesses to apply their creativity and innovation to solving sustainable development challenges” \cite{UN2015} (p. 29).
Being a recent phenomenon, research on the theme is at an embryonic state \cite{EMMA2021126781}. 
The first papers dealing with the relationships between firms and SDGs were primarily conceptual. For example, they suggested that managers integrate SDGs in their companies’ communication, turn them into actions (tactical level) and consider them in their strategies to contribute to the global challenges \cite{Redman2018}. Some other frameworks were developed to capture the alignment between existing activities and SDGs \cite{Tsalis2020}. However, to what extent companies contribute to SDGs is not a straightforward issue. Each activity that companies engage with has the potential to have both positive and negative effects on SDGs. When setting up a strategy to positively impact specific SDGs, companies need to recognise the diversity of consequences their strategies could have on other SDGs \cite{VanZantenvanTulder2021}. Scholars have advanced operational frameworks to support companies in understanding their impacts on the various SDGs dimensions. These push firms to evaluate their impacts not only in “core activities” but also in the other dimensions of the business \cite{VanZantenvanTulder2021empirical}. } \\
More broadly, empirical research is investigating the reasons why and factors that drive firms’ SDGs adoption \cite{Khaled2021, PATUELLI2022133723}, their challenges \cite{Dalton2020}, how SDGs are implemented in the firms’ strategies and activities \cite{Ike2019,Abd-Elrahman2020,Tabares2021} \aleX{ or taught in management education within business schools \cite{KOLB2017280}. A growing number of studies is focusing on how SDGs achievements are communicated \cite{garcia-sanchez_institutional_2020} and reported \cite{Dalton2020,Khaled2021,Silva2021}, also to map to what extent firms are contributing altogether to the global challenges \cite {Nylund2022,CALABRESE2022132324,Rosati2019,vanderwaal2020,Poddar2019,Silva2021}.
With this latter aim in mind, case studies \cite{Dalton2020,Ike2019,Tabares2021,PATUELLI2022133723, Lopez2020}, and the analysis of websites \cite{Lopez2020} can also be found.
The attempts to map companies’ SDG contributions mostly use companies’ reports. However, a series of caveats arise. First, (sustainability) reports are usually published once a year, making mapping companies’ engagement with SDGs somewhat delayed compared to the periods when the contributions took place. Second, the small dimensions of samples can capture only a tiny proportion of the firms’ contributions to SDGs. 
For example, Ref. \cite{Rosati2019} mapped firms’ SDGs contributions and the factors that explain firms’ involvement in SDGs based on their reports. As the authors only considered the companies’ 2016 sustainability reports, the study provides a valuable but very early analysis of the phenomenon, as SDGs were established in September 2015. Also, the paper has a limited sample size, considering 408 organisations. Similarly, Ref. \cite{vanderwaal2020} investigate the SDGs adoption two years after their introduction, examining the sustainability reports of the 2000 largest stock-listed businesses worldwide, while Ref. \cite{CALABRESE2022132324} analysed a sample of 385 sustainability reports disclosed by 235 companies in the period 2016–2020. Overall, the three studies agree that overall SDG involvement is quite limited, but this could also be a consequence of the early timing of the studies’ settings. 
Using empirical data about the CSR activities of the top 500 companies listed in the BSE (the largest stock exchange in India) between 2014-2016, Ref. \cite{Poddar2019} find that companies generally tend to carry out CSR activities aligned with SDGs from the social dimension. \aleY{These results seem to contradict the part of the literature arguing that  social sustainability has been overlooked across businesses and organizations, although it deserves more attention in practice and academic studies\cite{Prieto2022,doi:10.1080/13504509.2017.1408714}}. Differently, Ref. \cite{Nylund2022} find that companies from the Fortune’s Change the World 2019 list concentrate on environmental and social areas. However, the sample they base on is quite limited, as their empirical analysis is founded on 50 companies and 40 usable reporting documents. Interestingly enough, based on a sample of 385 reports from 235 firms from all over the world, Ref. \cite{CALABRESE2022132324} showed that firms in Healthcare contribute more than other sectors to the SDGs.
Though starting in the early years of SDGs adoption, Ref. \cite{Elalfy2020} take a much broader approach, analysing 14,308 reports from 9,397 organisations between 2016 and 2017 from the GRI dataset. }
The first findings show that businesses seem to focus on specific SDGs covering the three dimensions of sustainability. The most common ones seem to be SDG 3 (good health and well-being), 4 (quality education), 9 (industry, innovation and infrastructure) and 12 (responsible consumption and production) \cite{Elalfy2020}. 
\aleX{With a focus on entrepreneurship, Ref. \cite{Horne2020} innovates as far as the method, carrying out a semi-automated content analysis on web data on 588 new ventures in Germany in order to understand what role entrepreneurship plays in achieving the SDGs, finding that most activities align to social and economic goals, Again, though the method is innovative, the sample they base on is limited.}
As for now, research is overlooking the role of online social networks in understanding firms’ engagement with SDGs. To the best of our knowledge, only a few papers explore to what extent firms discuss SDGs themes on online social networks. \aleX{They take precise approaches, considering firms in the FinTech area \cite{Nicanor2021} or CEOs’ accounts \cite{Grover2019}.} \aleY{Regarding stakeholder engagement, while recent research on sustainability reporting finds a low level of stakeholder engagement disclosures 
\cite{Ardiana2023344 } }, studies \aleY{based on online social networks} highlight that  stakeholders \aleY{variously} interact on these themes \cite{Deluca2022,Zhou2022,Mehmood2022}. 
As for now, it seems that SDGs have limited relevance in the online debate \cite{Nicanor2021}. Also, there seems to be a limited involvement of stakeholders on SDGs posts, consistently with similar studies on CSR \cite{Manetti2016}. However, present studies use limited samples or focus on specific industries. 
\aleX{Considering the existing studies on the topic, we believe online social networks are promising tools to map companies’ engagement with SDGs in real-time. We aim to contribute to this issue by investigating to what extent firms are engaging with SDGs themes on Twitter. }

\begin{figure}[hbt!]
\centering
    \includegraphics[width=.6\columnwidth]{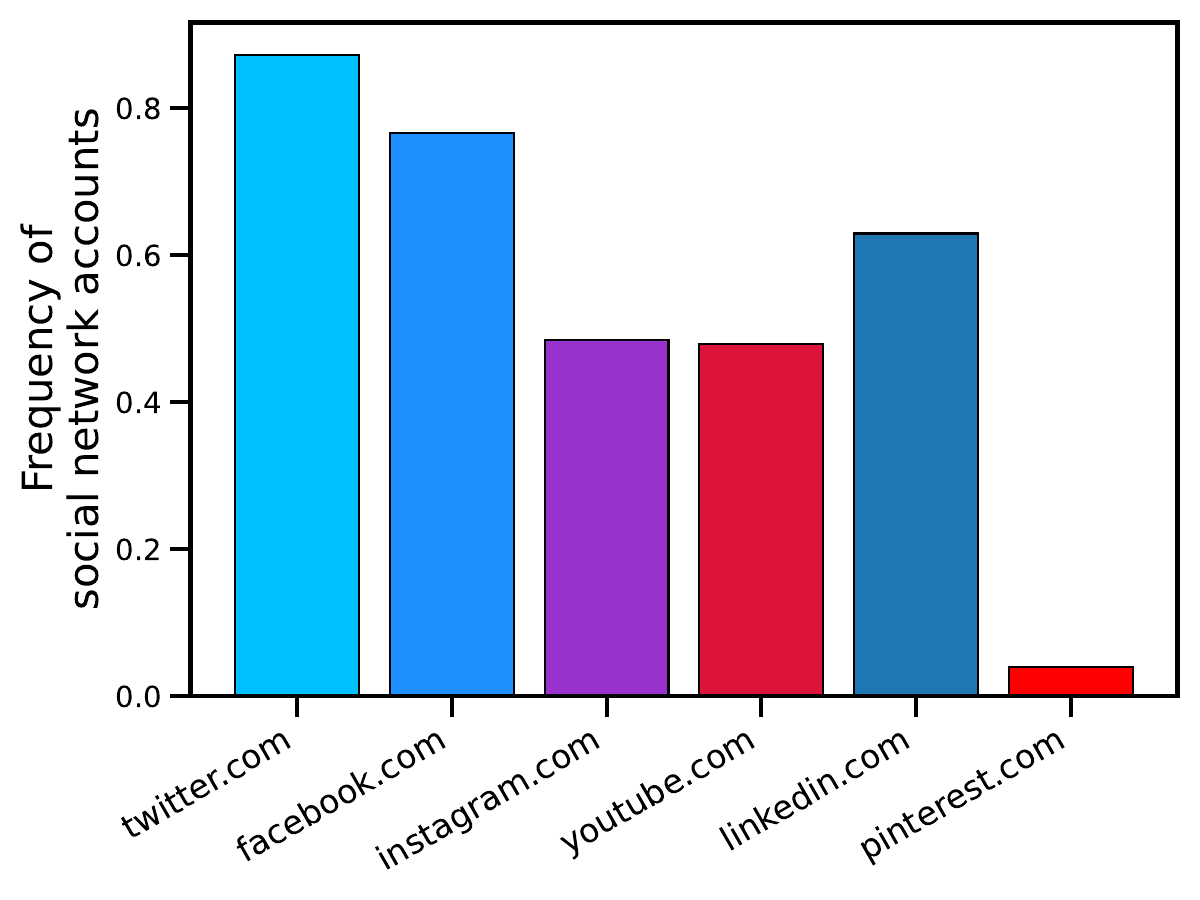}
    \caption{\textbf{Online social networks adoption \ale{among large firms in the UK.}} Twitter is the most popular online social network for large firms in the UK, overcoming the other ones.}
    \label{fig:twitter_UK}
\end{figure}

\section*{Results}
\begin{figure}[htb!]
\includegraphics[width=\linewidth]{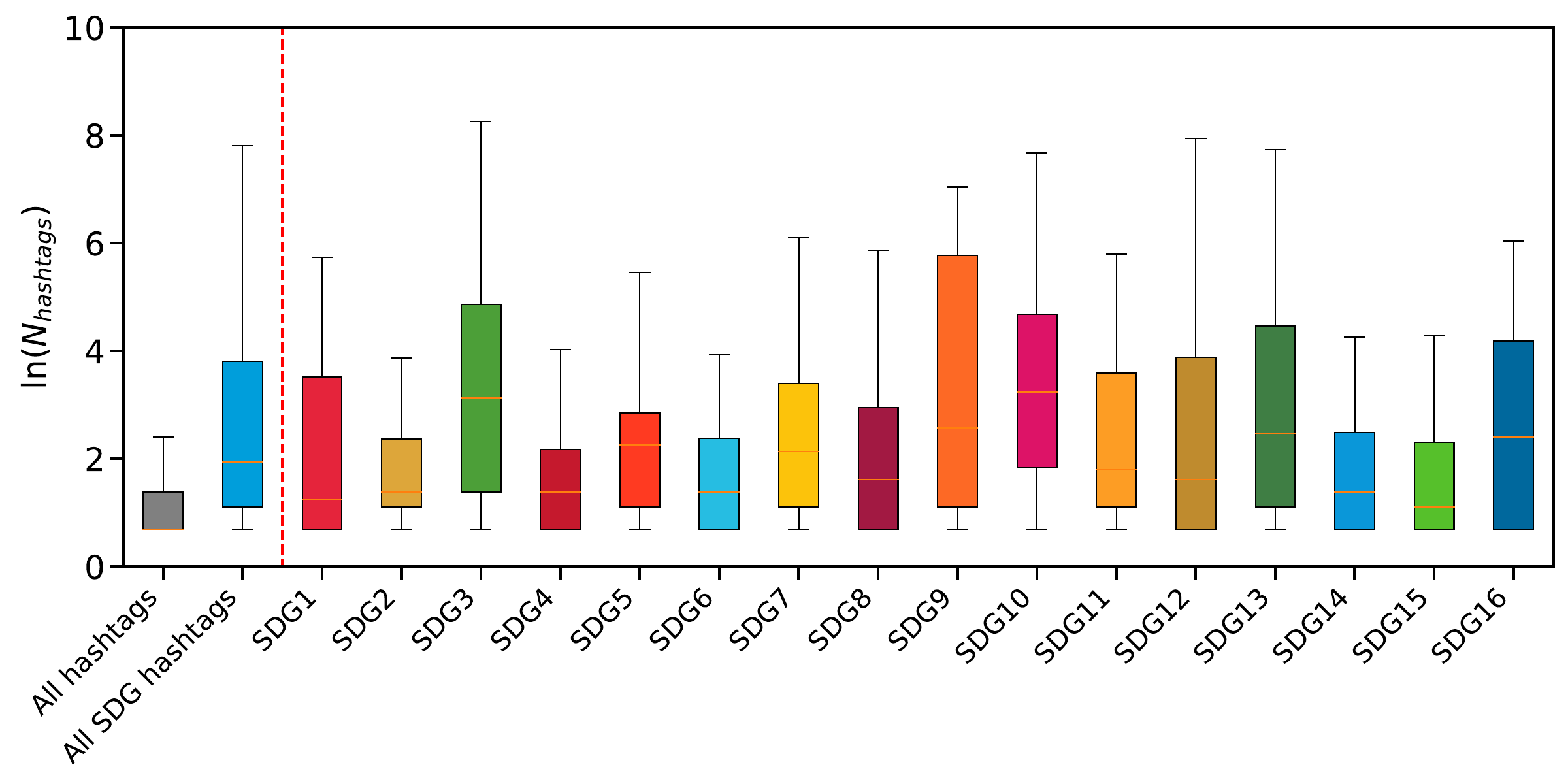}
\caption{\fab{\textbf{Distribution of the number of messages per SDG hashtag.} The boxplots compare the distribution of the number of messages in which each  hashtag appears for all hashtags (the grey box on the left) and SDG hashtags (all the boxes beyond the red line. Boxes are colored using the official indication from the United Nations, \url{https://www.un.org/sustainabledevelopment/wp-content/uploads/2019/01/SDG_Guidelines_AUG_2019_Final.pdf}) and; for all the SDGs hashtags (the sky blue box on the left). The boxplots show the distribution of the logarithm of the number of messages per hashtags, since the distributions are heavy-tailed. In this sense, boxplots may not be the perfect tool for capturing the distribution properties but can effectively deliver the message about the rough differences among the various distributions. In particular, SDGs hashtags appear more frequently than ``standard" hashtags in the communication strategy of large firms, thus representing \ale{crucial} topics.}}\label{fig:sdg_presence}
\end{figure}

\ale{\subsection*{Data description}
In order to highlight firms' \aleX{engagement with} SDGs, we first downloaded from Orbis (Bureau Van Dijk, https://www.bvdinfo.com/) the primary information regarding large companies (i.e. those with more than 250 employees), such as the name, address, number of employees, total assets,  NACE code, and the website. Then, we automatically extracted the Twitter account of the related firm from each website, if present. Please find more details about the automatic Twitter account search in the Methods section. 
We found that Twitter is an excellent tool for this analysis, as it is the most widely adopted online social network by UK large firms. As Fig.~\ref{fig:twitter_UK} shows, nearly 87.3\% of the largest UK firms have a Twitter account, overcoming other online social platforms.

Finally, we downloaded the timeline of each Twitter account using the official API (specifically, using the command: \href{https://developer.twitter.com/en/docs/twitter-api/tweets/timelines/api-reference/get-users-id-tweets}{\texttt{GET /2/users/:id/tweets}}) via \href{https://docs.tweepy.org/en/v3.5.0/index.html}{\texttt{tweepy}} python wrapper.
Doing so allows us to access the most recent $\sim3200$ messages. We focused on the period between 2021/02/17 and 2022/02/17. As in Ref.~\cite{Patuelli2021}, we consider active accounts in the entire period (and not just a portion). While this choice may appear too conservative, it allows concentrating our efforts on subjects that have continuously contributed to creating a shared narrative. Using this time restriction, we ended up with 3.1M tweets, out of which 596k retweets and 609k replies. 
As we focus on the interaction between Twitter accounts and hashtags, we excluded 318 accounts out of 6179 of the original dataset since they did not use any hashtag in the considered period.

}
\fab{\aleX{As a first step, we measured} the recurrence of SDG hashtags in our dataset, i.e. count how many messages contain hashtags related to SDGs, see Fig.~\ref{fig:sdg_presence}. \ale{SDGs are a crucial  topic in firms' communication: each SDG hashtag appears, on average, in 99.01 messages, against the 7.56 of the average hashtag in our dataset. }

}

\begin{figure}[htb!]
\includegraphics[width=.9\linewidth]{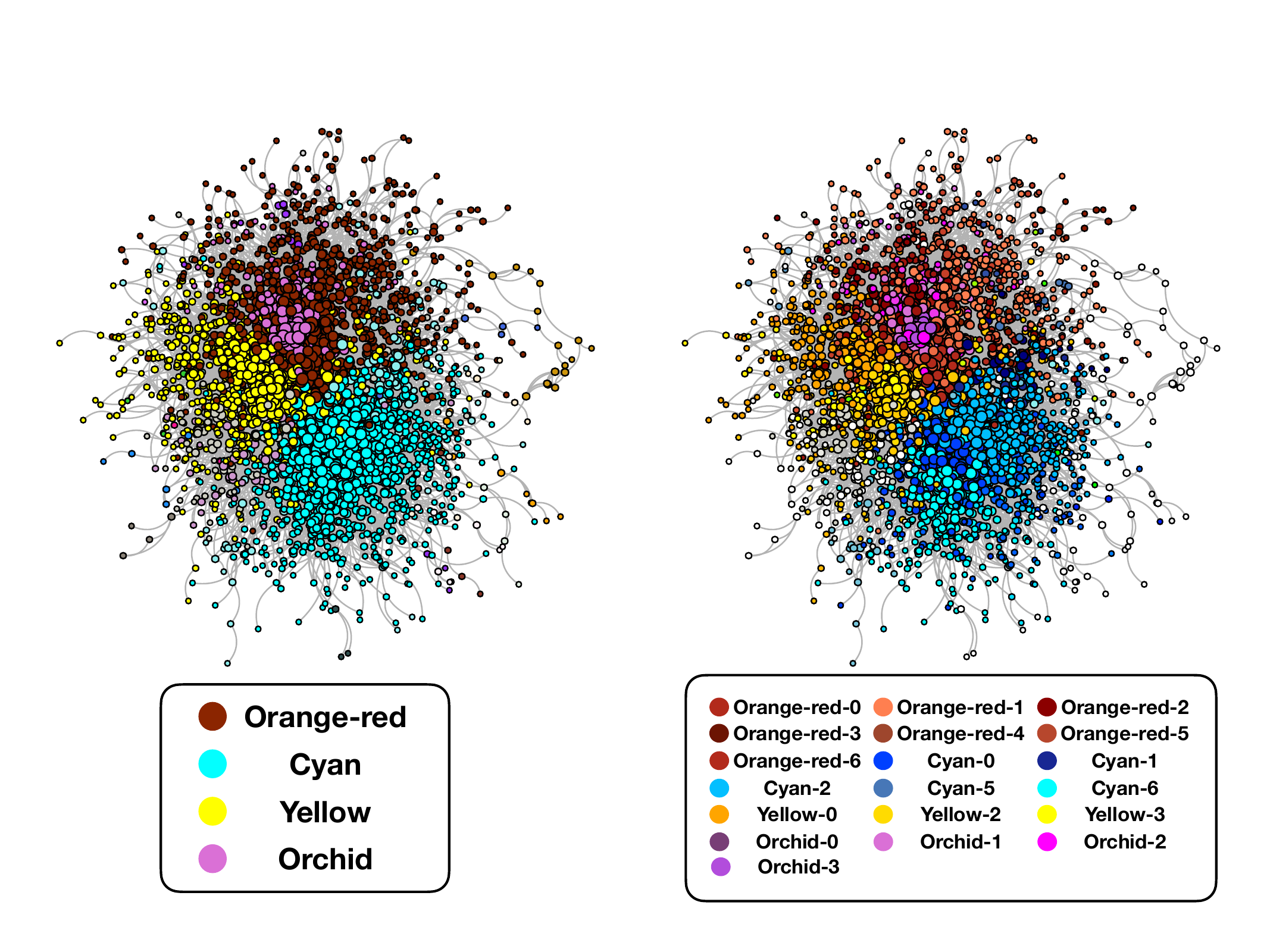}
\caption{\textbf{The Largest Connected Component of the validated projected network of users}. The dimension of the nodes is proportional to the logarithm of their degree. In the left panel, the different node colors represent the various communities, as detected by Louvain algorithm; in the right panel the colors identify the subcommunities (nodes that do not fall in one of the main subcommunities are plotted in white). \fabX{Please note that the contents of the various subcommunities are indicated in Tables~\ref{tab:cyan_hashtags},\ref{tab:orangered4_hashtags_1},\ref{tab:orangered4_hashtags_2}, \ref{tab:yellow1_hashtags} and \ref{tab:orchid_hashtags}.}}\label{fig:valnetuser}
\end{figure}

\ale{\subsection*{The validated network of Twitter accounts}}
\ale{To study how UK companies contribute to the evolution of common narratives, we considered the bipartite network composed of the firms’ Twitter accounts and their hashtags. In a bipartite network, nodes are divided into two sets, called \emph{layers}, and links can only connect nodes from different layers. \ale{In our application, the two layers include 5,859 accounts and 136,504 different hashtags.}
To highlight those accounts using the same hashtags, as in Ref.~\cite{Patuelli2021}, we use the validation projection procedure proposed in Ref.~\cite{Saracco2017}. In a nutshell, any couple of accounts are connected if the number of hashtags they used is statistically significant (i.e. it cannot be explained by their hashtag usage and the popularity of the various hashtags). Please find more details in the Methods section. The result of this validation projection is a monopartite network of Twitter 3,629 accounts and 59,158 links. The relative Largest Connected Component (LCC) is represented in Fig.~\ref{fig:valnetuser}. }\\

\ale{Before describing the network and its structure, we highlight a few remarks. First, the percentage of validated UK Twitter account nodes is 61.9\%. This percentage indicates the fraction of accounts whose usage of hashtags differs substantially from a random behaviour and whose communication strategy presents significant similarities \fab{with other accounts}. In this sense, a low frequency of validated nodes can mean that many accounts focus on the peculiarities of their communication. In contrast, a high frequency of validated nodes can mean that the communication is more homogeneous and strongly related to common narratives. \aleX{Instead, the percentage of validated large Italian firms' Twitter accounts was only 19.2\%.} \\
Second, SDGs are among the subjects contributing the most to developing common narratives. Validated users (those passing the validation procedure described above) contribute with no less than 85\% of the SDGs hashtags of the entire dataset; see Table~\ref{tab:sdg_usage_valusers}. Otherwise stated, most Twitter accounts using SDGs in their communications pass our filter. This result is remarkable since the validation procedure of Ref.~\cite{Saracco2017} is \fab{restrictive}, as tested in different contexts: such a strong signal indicates a non-trivial activity on SDG communication.}

\begin{table}[ht!]
\centering
\begin{tabular}{llllllll}
\hline\hline
 SDG01  & SDG02  & SDG03  & SDG04  & SDG05  & SDG06  & SDG07  & SDG08  \\
 \hline
 85.19\% & 97.91\% & 96.15\% & 92.65\% & 91.07\% & 90.54\% & 93.56\% & 92.50\% \\ \hline \\
 \hline\hline
 SDG09  & SDG10  & SDG11  & SDG12  & SDG13  & SDG14  & SDG15  & SDG16  \\
 \hline
 95.93\% & 94.89\% & 88.08\% & 96.13\% & 95.28\% & 92.80\% & 94.79\% & 86.55\% \\
\hline
\end{tabular}
\caption{\ale{\textbf{SDG hashtags used by validated users over their usage in the entire dataset.} These high percentages indicate that most active users using SDG hashtags are validated by the filtering procedure. In turn, it implies that SDGs are among the main subjects shaping the various common narratives of large firm's accounts on Twitter.}}
    \label{tab:sdg_usage_valusers}
\end{table}

\begin{figure}[htb!]
\includegraphics[width=\linewidth]{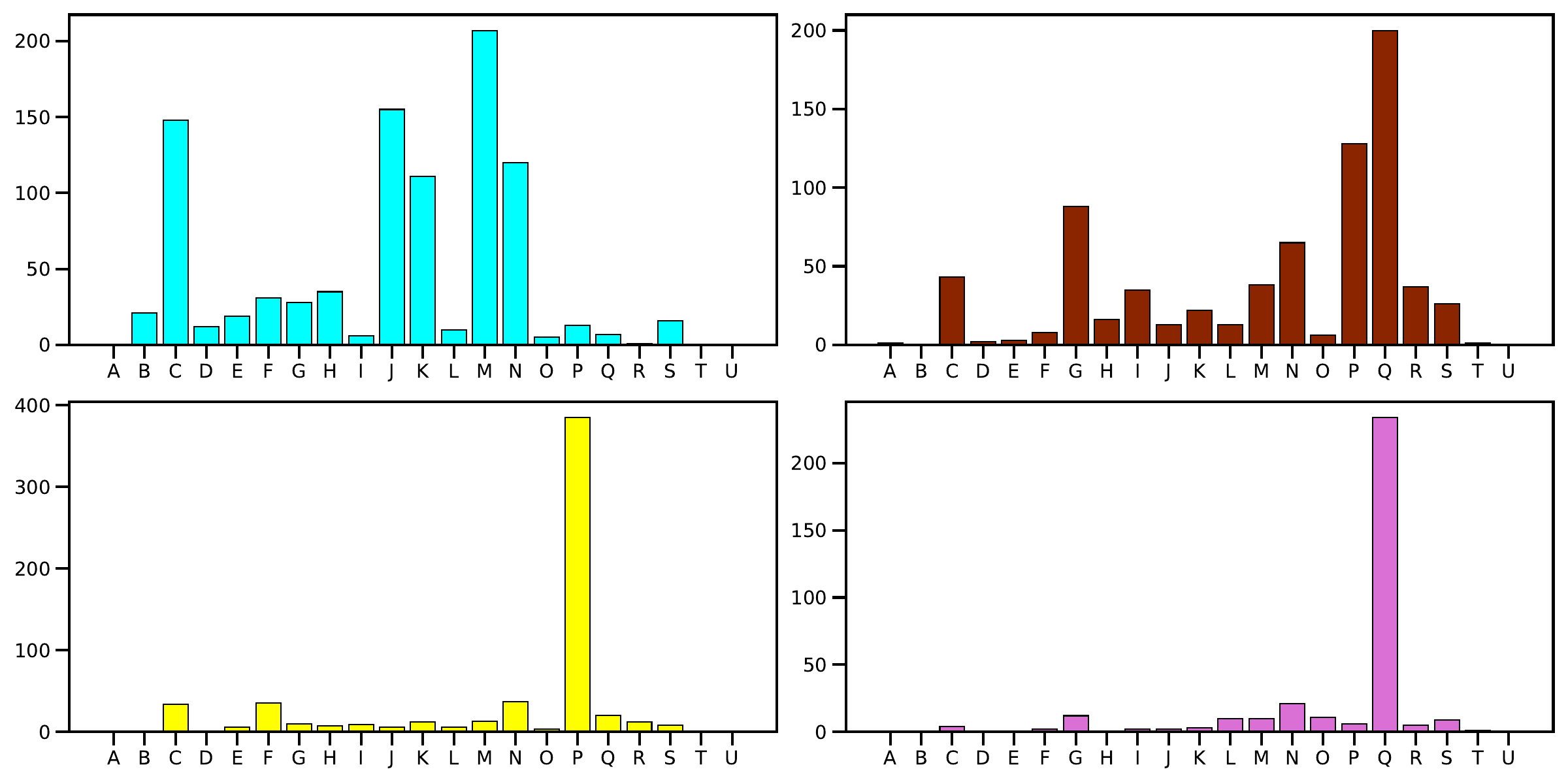}
\caption{\textbf{\aleY{The frequency of NACE Rev.2 (1 digit) sectors among the firms} of the four greatest communities in the validated network of Fig.~\ref{fig:valnetuser}}. \aleY{Please find the number of firms on the vertical axis; the} colors of the different bar charts are the same as in the left panel of Fig.~\ref{fig:valnetuser}. \ale{The identification of the various NACE is in Table~\ref{tab:nace} in the Methods section. To enhance clarity, we anticipate here the most present sectors, which are: C) Manufacturing; G) Wholesale and retail trade; repair of motor vehicles and motorcycles; J) Information and communication; K) Financial and insurance activities; M) Professional, scientific and technical activities; N) Administrative and support service activities; P) Education; Q) Human health and social work activities. 
} }\label{fig:valnetuser_nace}
\end{figure}

\begin{figure}[htb!]
\includegraphics[width=\linewidth]{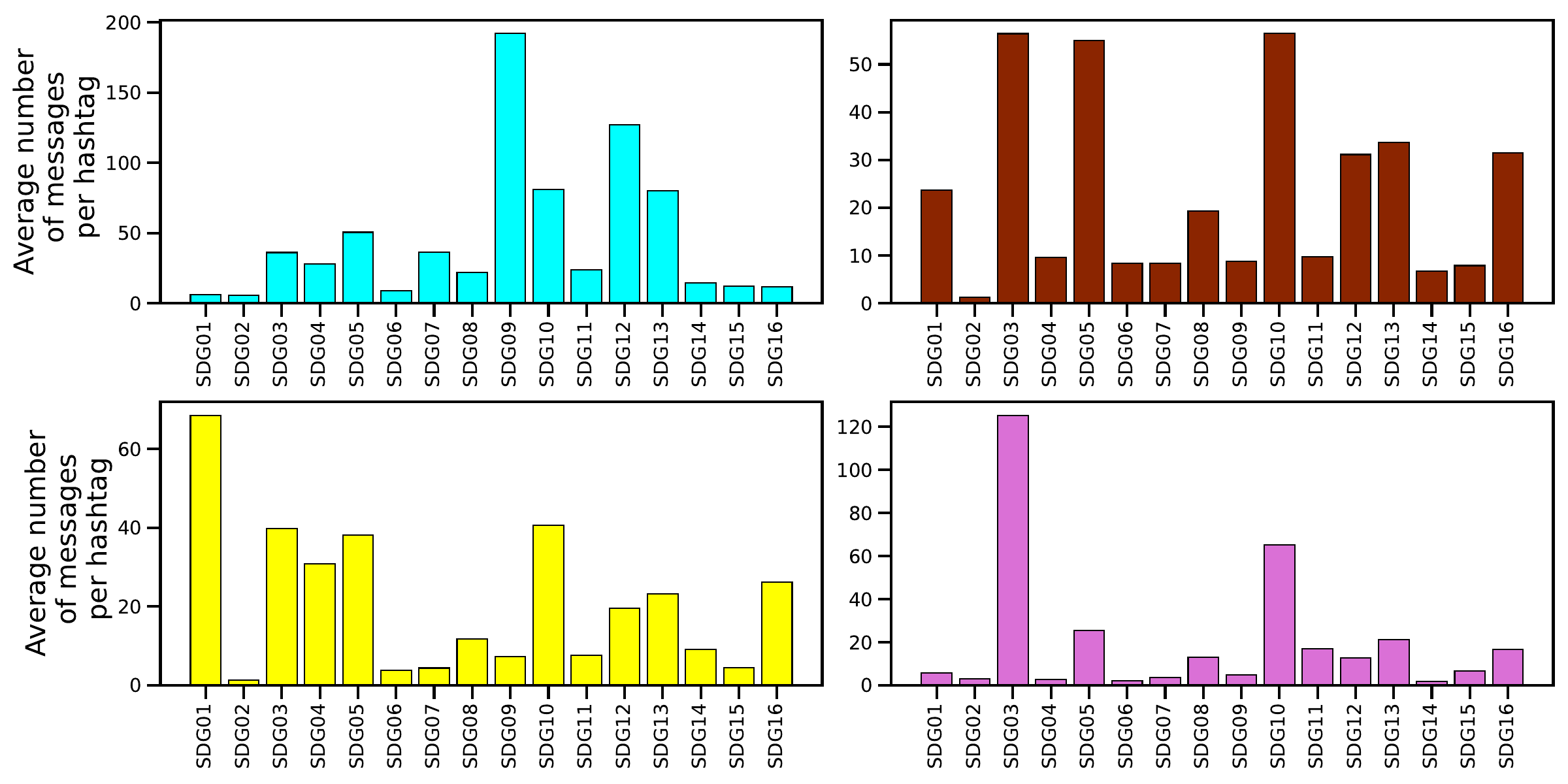}
\caption{\textbf{The SDG activity of the four greatest communities in the validated network of Fig.~\ref{fig:valnetuser}}. The colors of the different bar charts are the same as in the left panel of Fig.~\ref{fig:valnetuser}. The identification of the various hashtags with the different SDGs is described in detail in the Methods section.}\label{fig:valnetuser_sdg}
\end{figure}

\ale{\subsubsection*{Description of the communities \aleY{of Twitter accounts}}} 
To extract more information, we ran the Louvain community detection algorithm \cite{Blondel2008} on the validated network \aleY{of firms}, highlighting four main communities displayed in the left panel of Fig.~\ref{fig:valnetuser}. \aleY{The rationale is to find groups of firms contributing to the same common narrative, as captured by hashtags.} Rerunning the same Louvain community detection algorithm inside each community shows a more detailed description, which is represented in the right panel of Fig.~\ref{fig:valnetuser}.\\
The communities in Fig.~\ref{fig:valnetuser} mostly revolve around social themes, showing that CSR themes are indeed fundamental in firms' communication on Twitter, consistently with ~\cite{Patuelli2021} \aleY{while contradicting \cite{Prieto2022,doi:10.1080/13504509.2017.1408714}}.
Communities are generally coherent with the sector \aleY{(i.e. the economic activity)} the firms belong to, as captured by NACE code Rev.2 at 1 digit (see Fig.~\ref{fig:valnetuser_nace}; the description of the various codes can be found in Table~\ref{tab:nace}). 
This coherence reflects the themes discussed: the most addressed CSR themes are the ones closest to the firms’ sector, as represented in Fig.~\ref{fig:valnetuser_sdg}.\\ 
\ale{This confirms that CSR  changes according to the specific context \cite{Snider2003}. Moreover, we show that the social dimension appears more critical than the environmental one. Although this 
\fabX{result} contrasts with most previous literature \cite{Pedersen2010modelling}, it seems in line with more recent findings \cite{Patuelli2021}. }
Community Cyan is a sort of exception among the various groups, as it comprises three main sectors (Professional, scientific and technical activities; Information and communication; Manufacturing). Its top hashtags reflect digital innovation, environmental sustainability, social and economic themes (see Table~\ref{tab:cyan_hashtags}). They are coherent with the wide range of SDGs mentioned (SDG10 refers to the social dimension; SDG9 to the economic one; SDG12 and SDG13 to the environmental dimension). The other communities revolve around social themes and hashtags. Community Orange-red is composed of firms from two sectors (Human health and social work activities; Education) and focuses on social themes (\fab{see Tables~\ref{tab:orangered4_hashtags_1} and ~\ref{tab:orangered4_hashtags_2}}). Coherently, hashtags relate to SDGs from the social dimension, namely SDG3, SDG5 and  SDG10. Community Yellow comprises firms from one sector (Education) and mainly discusses social themes (see Table~\ref{tab:yellow1_hashtags}). The most mentioned SDGs come from the social dimension: SDG1, SDG3, SDG4, SDG5, SDG10, and SDG16.
Similarly, Orchid has companies from only one sector (Human health and social work activities). It is focused on social themes as well (see Table~\ref{tab:orchid_hashtags}). As community Yellow, its SDGs belong to the social dimension:  SDG3, SDG5, and SDG10.

%
%
%

\subsection*{\fab{Hashtag frequency}}
In this subsection, we will focus on the four major communities, i.e.  Cyan, Orange-red, Yellow and Orchid. For each community, we show the most recurring hashtags in the \ale{biggest subcommunities }(with more than 50 nodes), which reflect the main themes that businesses discuss. The following analysis is based on the results summarised in Tables~\ref{tab:cyan_hashtags},~\ref{tab:orangered4_hashtags_1},~\ref{tab:orangered4_hashtags_2},~\ref{tab:yellow1_hashtags} and ~\ref{tab:orchid_hashtags}.
\par
The Cyan subcommunity shows some diversity in its themes, which span from digital innovation and environmental sustainability to social and economic themes. While subcommunity n. 0 is focused on digital innovation themes (e.g., “cloud”, “technology”, “digital”, “digitaltransformation”, “innovation”, “webinar”), subcommunity n. 2 mostly focuses on environmental sustainability (e.g. “sustainability”, “netzero”, “climatechange”, “earthday”). While mostly focusing on social themes, subcommunity n. 1 also contains some environmental themes (e.g., “blackhistorymonth”, “pridemonth”, “internationalwomensday” for the social side; “earthday” and “sustainability” for the environmental one). Similarly, subcommunity n. 5 does not have a clear orientation towards one theme: it contains both environmental and economic themes (e.g., “sustainability”, “climatechange”, as well as “esg”, “inflation” and “supplychain”). Differently from the other subcommunities, n. 6 does not have a specific focus (it includes various hashtags, as “covid”, “budget”, “webinar”, “internationalwomensday”, “Brexit”, and “podcast”). 
\par
Compared to the Cyan community, the themes in the Orange-red, Yellow and Orchid show more homogeneity. The Orange-red community is highly focused on social themes. Among the seven subcommunities, they all show hashtags related to social themes. Six of them have social themes as the prevalent ones within the subcommunity. For example, subcommunity n. 1, which has a high prevalence of hashtags related to the social dimension, has “internationalwomensday”, “mentalhealthawarenessweek”, “pridemonth”, “blackhistorymonth” among the most frequent hashtags. Something similar happens with subcommunities n. 2, 3, 4 and 5. Conversely, the subcommunity n. 0, while having a few hashtags related to social themes, is more focused on festivities (e.g. “Christmas”, “valentinesday”, “halloween”). 
\par
Community Yellow discusses social themes in all three subcommunities, which are mostly homogeneous. For example, all three subcommunities mention “mentalhealthawarenessweek” and “internationalwomensday” among their top 10 hashtags, with some differences in other hashtags. Only subcommunity n. 3 \ale{comprises} themes related to engineering education (including hashtags like “engineering”, “education”, “construction”, “apprenticeship”). 
\par
Community Orchid is focused on social themes as well. It includes Covid themes among all its subcommunities: in this case, the social dimension is connected to the pandemic. For example, all four subcommunities associate “covid” and “covidvaccine” with “nhs”, “internationalwomensday”, and “mentallhealthawarenessweek”. Overall, the hashtag “covid” is generally found in many subcommunities 
, but it is not associated with other related words. It has a higher relevance only in the Orchid community. 

\ale{\subsection*{Stakeholder engagement}
This subsection focuses on stakeholder engagement with the narratives developed by firms’ accounts. 
First, the dataset shows that the average number of retweets and likes per hashtag is 15.85 and 23.16. This shows that UK firms’ stakeholders tend to use more likes than retweets when interacting on Twitter. However, further analyses reveal that this pattern is the opposite when stakeholders interact with companies on SDGs subjects. \ale{As Fig.~\ref{fig:stakeholders_all} shows, when stakeholders interact with SDGs hashtags, they put a lower number of likes but retweet more than they do with non-SDG hashtags. The average numbers of likes and retweets per SDG hashtag are 6.71 and 19.83, highlighting a higher engagement on SDGs themes.}
We also highlight that stakeholder engagement with large companies in the UK is different compared to Italy~\cite{Patuelli2021}, where the average number of retweets and likes per hashtag were 5.39 and 14.83.\\

\begin{figure}[ht!]
    \includegraphics[width=\columnwidth]{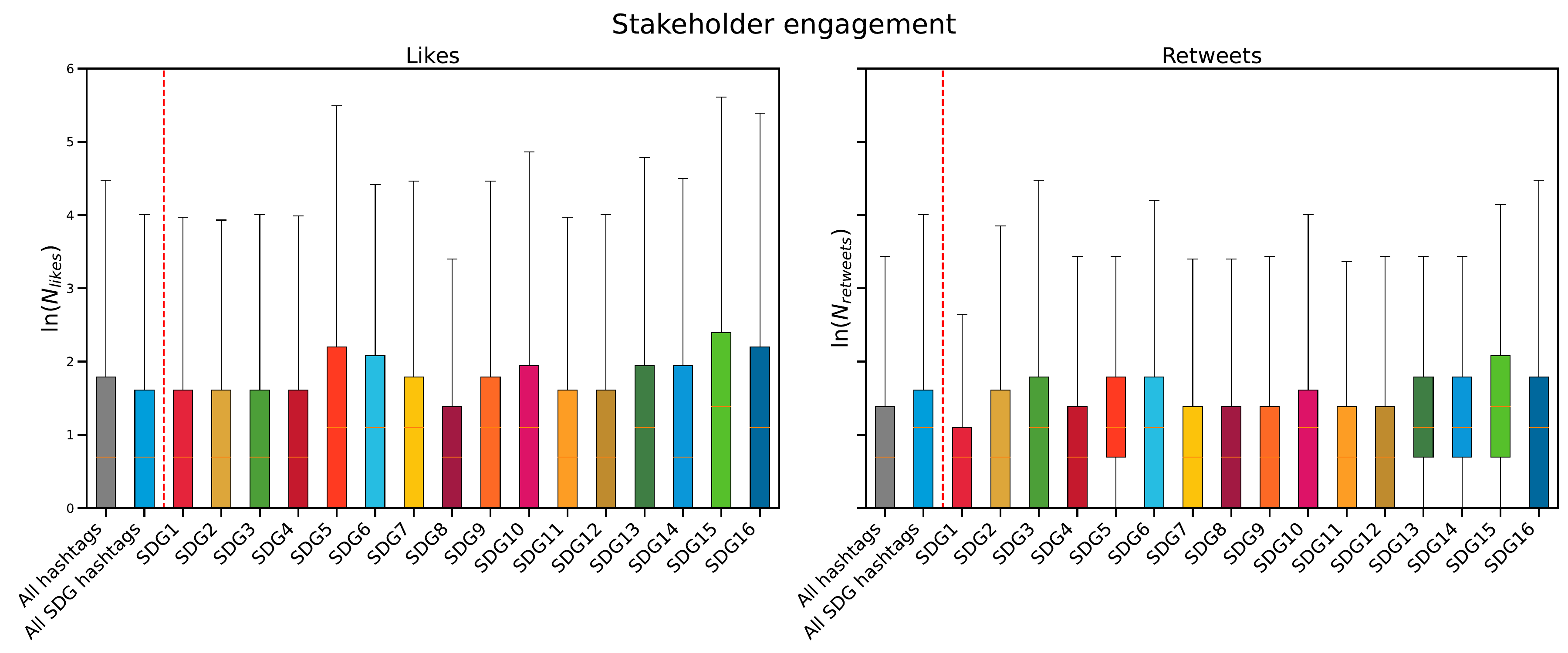}
    \caption{\ale{\textbf{Stakeholder engagement on the SDGs.} These boxplots compare the distribution of the number of likes and retweets for all hashtags (the gray box on the left), for each SDG (all the boxes beyond the red line; boxes are colored using the official indication from UN, \url{https://www.un.org/sustainabledevelopment/wp-content/uploads/2019/01/SDG_Guidelines_AUG_2019_Final.pdf} ), and for all the SDGs hashtags (the sky blue box on the left). The boxplots show the distribution of the logarithm of the number of likes and retweets. We used the logarithms because the distributions are heavy-tailed. In this sense, boxplots may not be the perfect tool for capturing the distribution properties but can effectively deliver the message about the rough differences among the various distributions.}} 
    \label{fig:stakeholders_all}
\end{figure}

Stakeholder engagement on the various SDGs depends on the community and sector. For example, in community Orange-red, SDG5 and SDG16 hashtags (i.e. `Gender Equality’ and `Peace, Justice, and Strong Institutions’) received more retweets, on average, than other SDG hashtags, and more than random hashtags. This community mainly comprises firms in sectors `P’ and `Q’ (`Education’ and `Human health and social work activities’). Analogous considerations can be done for the other communities. 
\aleY{Doing so, our findings seem to contradict \cite{Ardiana2023344,Manetti2016} and integrate \cite{Deluca2022,Zhou2022,Mehmood2022}, showing a higher involvement of stakeholders in SDGs themes compared to non-SDGs related Twitter posts. }

}





 

\section*{Concluding remarks}


This paper presents large UK firms’ discussions on Twitter, specifically focusing on SDGs. It shows that: 1) SDGs are the themes that unite firms’ discussions; 2) the social dimension is prevalent, compared to the environmental and economic ones; 3) the interest in specific SDGs depends on the community and sector a firm belongs to; 4) stakeholders are highly engaged on SDGs themes, using more retweets than likes when interacting with a tweet that contains an SDG-related hashtag; 5) overall, large UK firms and stakeholders show substantially different behaviours compared to the Italian ones. We will discuss these points in the following paragraph.\\ 
First, communities of discussion naturally arise from the data. These communities are uniform and based on common narratives. Most importantly, the shared narratives are centred around SDG themes. Large UK firms use Twitter to participate in broader discussions on widely acknowledged themes (such as “internationalwomensday”). Thus, we believe that our results for the UK support stakeholder theory: large firms use Twitter to engage in discussions on highly socially relevant themes. This finding gives a different perspective compared to previous research, which states that CSR themes are overlooked by firms in their communications on online social networks \cite{Manetti2016,Gomez2016,Etter2013,Patuelli2021} \aleX{and that companies are scarcely involved in the SDGs \cite{Rosati2019,vanderwaal2020,CALABRESE2022132324}}.
 While not contradicting these previous studies, we show that SDG themes unify the firms’ discussions, creating different communities \fab{in the UK debate on Twitter}. In doing so, \aleX{we show that SDGs themes indeed entered the firms’ communication level \cite{Redman2018}}. We also highlight the importance of integrating different methodologies into business research, uncovering patterns that would not show using traditional methods \cite{choudhury2018machine}. \\
Second, the recurrent themes in the communities mainly focus on the social dimension, with discussions on environmental and economic themes that are present but less relevant. Our findings oppose traditional CSR literature, which maintains that environmental themes are the primary dimension \cite{Pedersen2010modelling}. \aleX{However, they seem consistent with more recent studies on SDGs, finding either that companies are focused on the social dimension \cite{Poddar2019} or that they do not concentrate on a single dimension only \cite{Nylund2022,CALABRESE2022132324,Elalfy2020, Horne2020}}.\\  
Third, we highlight that the interest in SDGs depends on the community a firm belongs to, and the community mostly depends on the firm’s sector. This is consistent with previous works about SDGs, which argue that the interest in SDGs depends on the sector the firm belongs to \cite{Elalfy2020, CALABRESE2022132324}. It is also consistent with previous findings about large Italian firms discussing CSR themes\cite{Patuelli2021}. Both in the UK and Italy, large firms’ dialogue largely depends on the sectors to which they belong. Only community Cyan discusses the themes shared by large Italian firms \cite{Patuelli2021} (i.e. the digital transformation, environmental sustainability, Covid and the economic dimension). \aleX{Thus, UK firms’ behaviour appears to be substantially different from the one of large Italian firms, as described in Ref. \cite{Patuelli2021}}. \\ 
Fourth, results highlight stakeholder engagement with retweets is higher on SDG-related tweets than on general tweets. As retweets are a more significant endorsement of the author of the post~\cite{Conover2011,Conover2011a, Conover2012}, the higher number of retweets on SDGs themes highlights a more significant engagement with the global challenges.
\aleX{Overall, our results provide a map of large UK firms’ engagement with themes related to SDGs. The results also show} a different use of Twitter by UK firms compared to the Italian ones. Thus, we highlight that consistently with institutional theory \cite{Hofstede1984}, different institutional and cultural settings translate into different behaviours, including corporate communications on online social networks. These differences are not limited to localized behaviours. Instead, they appear to relate to the fundamental reasons why companies interact in online social networks, highlighting that the results support different theories in the two countries.\\

\section*{Contributions, practical implications and future research paths}

Our paper brings several contributions. 
First, we contribute to institutional \cite{Hofstede1984}, stakeholder and legitimacy theories \cite{Brown1998, Guthrie1989}, \aleX{explaining UK firms’ attitudes on Twitter and comparing them with Italian firms’}. We argue that stakeholder and legitimacy theories coexist and explain firms’ behaviours on Twitter in the two countries. Following the institutional theory, we believe that different institutional settings, values and cultures explain these different behaviours and their reasons.\\ 
Second, we answer previous calls to map firms’ contributions to the SDGs \cite{Mio20203220,deVilliers2021598}. \aleX{As research about firms and SDGs is still at an embryonic state \cite{EMMA2021126781}, we contribute to} advance preliminary studies \cite{Deluca2022} with an interdisciplinary approach on a wide dataset, \aleX{also providing a novel set of keywords to detect firms' engagement with SDGs on Twitter}. Our research highlights that Twitter posts concerning SDGs themes unify the firms, naturally creating discussion communities. It also shows the prevalence of the social dimension, as opposed to the environmental and economic one, and a higher engagement of stakeholders on these themes compared to general posts.\\
\aleX{
This paper brings several practical implications which might be relevant for firms, policymakers and in management education. In short, it provides a novel tool to monitor the influence of the private sector on the implementation of the 2030 Agenda. Knowing what businesses are doing and how they are engaging in global challenges in real time is essential to handle any issues promptly and effectively. On the one hand, managers could use online social network data to understand what their competitors and firms from other sectors contribute to regarding SDGs. On the other hand, policymakers could gain an advantage from the timeliness of using big data to capture firms' engagement in sustainable development objectives. Compared to the analysis of corporate reports, widely used in the academic literature, our research proposes a novel and nearly real-time approach to monitoring firms' engagement with the SDGs. Plus, it can capture a high number of firms at the same time. Thus, our approach could serve as an additional tool for policymakers to monitor SDG progress, with a chance to develop real-time policies to improve the firms' engagement with global objectives.
Additionally, this research might be of interest to business schools. As proposed by Ref. \cite{KOLB2017280}, business schools must educate future managers with a holistic perspective, integrating sustainable management education with interdisciplinary approaches. Our paper could serve as a tool to increase awareness of the different methodologies a (sustainability) manager can use to understand the environment. } 
\\
\aleX{ This paper has several limitations that open new paths for future research. The first limitation is the time frame considered, which is a specific year (from 2021/02/17 to 2022/02/17). As SDGs are an evolving phenomenon,} it would be interesting to go back in time and check if and how this trend increased in the past years. Also, according to the UN, SDGs should be reached by 2030. \aleX{A future study could investigate how companies’ engagement with SDGs had changed over} the whole 15-years period.

\aleX{The second limitation is geographical. Our results are based on one country, the UK. They are consistent with Ref. \cite{Patuelli2021}, but contradict Ref. \cite{Pedersen2010modelling}. Thus, it would be interesting to check if the higher interest we found in social themes rather than environmental ones holds} for other contexts. Future research should investigate to what extent firms discuss the social and environmental dimensions on online social networks on a broader scale. \aleX{This kind of research could substantially contribute to the academic literature focused on the perceptions of the responsibilities of businesses towards society \cite{Pedersen2010modelling}.}\\
Third, while we can assume that online communication reflects the firms’ strategies and activities  \cite{Pilar2019}, we do not have enough data to claim that companies are actually pursuing the SDGs they are discussing in online social networks. Further research could dig deeper into this issue to unravel how firms’ communication about SDG themes is consistent with their real-world activities on global challenges.
\aleX{Fourth, we acknowledge that what companies communicate on Twitter does not represent the whole picture of their contributions to SDGs. A company might focus on sustainable activities in its core business while negatively impacting some SDGs with its ancillary activities \cite{VanZantenvanTulder2021,VanZantenvanTulder2021empirical}. While this paper aims at mapping companies’ engagement towards SDGs, a limitation is that our method tends to map the positive contributions companies declare on online social networks while not revealing the negative impacts that might arise. }
\aleY{Fifth, our paper considers a specific category of stakeholders, namely, online Twitter users. While our measure of stakeholder engagement is generally regarded as appropriate \cite{Manetti2016}, it should be noted that a firm's stakeholders are various and beyond online users. Stakeholders might engage in a business' activities offline. Thus, our measure of stakeholder engagement reflects only a part of the phenomenon.}
\aleX{Last, our research is only based on large firms and does not consider small and medium-sized enterprises’ (SMEs) contributions. Although large firms have a higher social impact \cite{Campopiano2015,Iaia2019} and engage more in sustainable activities compared to SMEs  \cite{giacomini2020environmental,CALABRESE2022132324}, SMEs represent the majority of firms in Europe. Future studies should try to capture SMEs’ engagement with SDGs, though facing the difficult challenge of data availability.}

\section*{Methods}



\subsection*{From websites to Twitter accounts}
\ale{As a first step, we downloaded companies' websites from Orbis and, automatically accessing them, we got the relative Twitter accounts, when present. 
In order to test our scraping algorithm, we took a sample of 100 websites and manually extracted the Twitter accounts from them.\\ 
Then, if both the human and the scraping algorithm agree on the account, we assign a true positive (TP) to each of them. If they cannot find any Twitter account, we assign a true negative (TN) to both. In other cases ( i.e., when one finds an account and the other does not, or if they disagree on the account found), we manually checked directly from Twitter which method returns the correct answer. Even if the scraping algorithm performances are not astonishing, they overcome the human ones. Please find the performances of the automated tool for getting the Twitter accounts in Table~\ref{tab:machine_vs_human}.
}

\begin{table}[htb!]
    \centering
    \begin{tabular}{|c||c|c|}
        \hline\hline
        & Machine & Human\\
        \hline\hline
        precision=$\frac{TP}{TP+FP}$ & \textbf{96.7\%} & 94.9\%\\
        \hline
        accuracy=$\frac{TP+TN}{P+N}$ & \textbf{83.0\%} & 81.0\%\\
        \hline
        sensibility=$\frac{TP}{TP+FN}$ & \textbf{79.5\%} & 77.8\%\\
        \hline
        specificity=$\frac{TN}{TN+FP}$ & \textbf{92.6\%} & 89.3\%\\
        \hline
    \end{tabular}
    \caption{\textbf{Scraping algorithm performances vs. human annotation; best performances in bold.} \ale{Machine performances always overcome human ones. From the data above, the primary human and machine problem seems to be missing the existing Twitter accounts (low sensibility). At the same time, if they find anything, they get the right one in most cases (high precision). The high specificity tells us that both the human and the machine can spot when the Twitter account is absent.} }
    \label{tab:machine_vs_human}
\end{table}

\subsection*{Hashtag data cleaning: edit distance}
In order to properly consider the issue of misspelt hashtags or to consider as a single word singular and plural nouns, we used edit distance\fab{, as implemented by the \href{https://github.com/anhaidgroup/py_stringmatching}{\texttt{py\_stringmatching}} python module~\cite{py_stringmatching}}. In order to obtain the most effective threshold, we randomly picked couples of keywords and selected the first 100 couples with an edit similarity score greater than 0.8. Then, we manually checked when the hashtags effectively represent different words or if they refer to the same concept. For this 100 couples' sample, we calculated the precision and the accuracy for the various values of the threshold (sensibility and specificity are trivial and do not carry any relevant information): the most effective threshold for edit similarity is 0.86.

\subsubsection*{Hashtag cleaning}
\ale{A significant part of hashtags refers to acronyms, and comparisons among them may cause false matches between unrelated terms. Thus, we first removed digits from the hashtags, except for `Euro2020', since it is an event that was central in the UK in the analysed period. Removing digits would have introduced mismatches and errors. Then we turned all hashtags to lower cases and considered their frequencies. In principle, using edit distance for hashtag cleaning, we should have compared all couple of hashtags, thus performing $O(N^2)$ tests.\\ 
In order to limit the efforts dedicated to hashtag cleaning to $O(N)$, we implemented the following procedure. 
First, we selected all hashtags appearing in the dataset more than 50 times, resulting in 922 different hashtags. In this `benchmark set', we first select all couples of words displaying an edit distance greater than the edit threshold of 0.86. \fab{Among those, we choose the less frequent hashtag for every couple} and remove it from the benchmark set, resulting in a total of 916 different hashtags. We finally compared all hashtags with the ones in the benchmark set. All hashtags that displayed an edit similarity greater than the threshold with another hashtag in the benchmark set were then substituted with \fab{their} more frequent partner. After the cleaning, we have 136,504 different hashtags.}

\subsection*{Bipartite Configuration Model analysis}
After the cleaning, we build a bipartite network in which the two layers represent respectively firms' Twitter accounts and the used hashtags, as in Ref.~\cite{Patuelli2021}. The two layers includes respectively 5,859 accounts and 136,504 different hashtags.\\ 
\fab{In order to have a proper benchmark for our analyses, we leverage on the Bipartite Configuration Model (BiCM,~\cite{Saracco2015}), i.e. the extension to bipartite networks of the entropy-based null-models reviewed in Ref.~\cite{Cimini2018}.\\ 
In a nutshell, the procedure is based on 3 main steps. First, we define an ensemble of (bipartite) networks, all having the same number of nodes per layer as in the real systems, but displaying all possible edge configurations, from the empty graph to the fully connected one. We then maximise the Shannon entropy associated to the ensemble, constraining some topological quantities of the network ~\cite{park2004statistical} (this approach replicate the approach of Jaynes for deriving Statistical Physics from Information theory~\cite{Jaynes1957a}). In particular in the Bipartite Configuration Model, we constrain the average (over the ensemble) degree sequences for both layers to the values observed in the real system. Finally, in order to obtain the numerical value of the related Lagrangian multipliers, we maximize the Likelihood of the real system, i.e. the probability, according to our null-model, of getting the observed network~\cite{Garlaschelli2008}.\\
Using the present procedure, we are getting a benchmark that is maximally random (due to the entropy maximization), but still tailored on the real system (due to fixing the degree sequences to one observed in the real network). In the following, we will first introduce briefly the formalism, then the Bipartite Configuration Model and, finally, its application for the validation of the co-occurrences.


\subsubsection*{Formalism}
Let us call $\top$ and $\bot$ the two layers of the bipartite network and use Latin and Greek indices to indicate elements in the respective sets; we indicate with $N_\top$ and $N_\bot$, respectively, the dimension of the two layers. The biadjacency matrix $\mathbf{B}$ associated to the bipartite network is a $N_\top\times N_\bot$ matrix whose generic entry $b_{i\alpha}$ is either 1 or 0 if either there is or there is not a link connecting node $i$ with node $\alpha$. Therefore the degree sequences for both layers read $k_i=\sum_\alpha b_{i\alpha}\,\forall i\in N_\top$ and $h_\alpha=\sum_i b_{i\alpha}\,\forall \alpha\in N_\bot$.

\subsection*{The Bipartite Configuration Model}
Let us call $\mathcal{G}_\text{Bi}$ the bipartite networks' ensemble containing all possible graphs in which the dimension of the layers are respectively $N_\top$ and $N_\bot$. If $S=-\sum_{G_\text{Bi}\in\mathcal{G}_\text{Bi}} P(G_\text{Bi})\ln P(G_\text{Bi})$ is the Shannon entropy, its maximization, constraining the average degree sequence on both layers, is equivalent to the maximization of $S'$ defined as
\begin{equation*}
    S'=S+\sum_i\eta_i \Big[k_i^*-\sum_{G_\text{Bi}\in\mathcal{G}_\text{Bi}} P(G_\text{Bi})k_i(G_\text{Bi})\Big]+\sum_\alpha\theta_\alpha \Big[h_\alpha^*-\sum_{G_\text{Bi}\in\mathcal{G}_\text{Bi}} P(G_\text{Bi})h_\alpha(G_\text{Bi})\Big]+\zeta\Big[1-\sum_{G_\text{Bi}\in\mathcal{G}_\text{Bi}} P(G_\text{Bi})\Big], 
\end{equation*}
where quantities with an asterisk $*$ represent the values observed in the real network and $\eta_i,\,\theta_\alpha$ and $\zeta$ are the Lagrangian multipliers associated, respectively, to the degree sequence of layer $\top$, to the degree sequence of layer $\bot$ and to the normalization of the probability $P(G_\text{Bi})$. The maximization of the $S'$ returns the functional form of the probability per graph $P(G_\text{Bi})$ in terms of the Lagrangian multipliers:
\begin{equation*}
    P(G_\text{Bi})=\prod_{i,\alpha}\dfrac{e^{-(\eta_i+\theta_\alpha)b_{i\alpha}(G_\text{Bi})}}{1+e^{-(\eta_i+\theta_\alpha)}}.
\end{equation*}
Therefore, $P(G_\text{Bi})$ can be interpreted as the product of independent probability $p_{i\alpha}=\frac{e^{-(\eta_i+\theta_\alpha)}}{1+e^{-(\eta_i+\theta_\alpha)}}$ of connecting node $i$ with node $\alpha$. In order to get the numerical values of Lagrangian multipliers $\eta_i$ and $\theta_\alpha$, we can maximise the Likelihood associated to the observed network: it can be shown (see Ref.~\cite{Garlaschelli2008}) that it is equivalent to setting
\begin{equation*}
    \left\{
    \begin{array}{l}
    k_i^*=\langle k_i\rangle=\sum_\alpha p_{i\alpha}=\sum_\alpha \dfrac{e^{-(\eta_i+\theta_\alpha)}}{1+e^{-(\eta_i+\theta_\alpha)}};\\
    h_\alpha^*=\langle h_\alpha\rangle=\sum_i p_{i\alpha}=\sum_i \dfrac{e^{-(\eta_i+\theta_\alpha)}}{1+e^{-(\eta_i+\theta_\alpha)}}.
    \end{array}
    \right.
\end{equation*}

\subsection*{Validated projection of bipartite networks}
\ale{Using the Configuration model defined in the previous subsection, it is possible to validate the projection of the bipartite network on one of the two layers. This procedure aims at stating the statistical significance of the co-occurrences observed in real systems. Consider, for instance a couple of nodes $(i,j)$ belonging to $\top$ layer: the probability that they both link node $\alpha\in\bot$ is }
\begin{equation}\label{eq:p_v}
    P(V^{ij}_\alpha=b_{i\alpha}b_{j\alpha})=p_{i\alpha}p_{j\alpha}, 
\end{equation}
where $V^{ij}_\alpha$ is the event ``both $i$ and $j$  are linked to $\alpha$'' and $p_{i\alpha}$ is the probability of connecting nodes $i$ and $\alpha$. Using Eq.~\ref{eq:p_v}, we can calculate the probability that the total number of co-occurrences between $i$ and $j$ is exactly $n$ as the sum of the contributions from all possible ways to choose $n$ nodes in $\bot$ layer. If we call $A_n$ this last quantity, the probability of observing $V_{ij}=\sum_\alpha V^{ij}_\alpha=n$ is

\begin{equation}\label{eq:PB}
f_{PB}(V_{ij}=n)=\sum_{A_n}\left[\prod_{\alpha\in A_n}p_{i\alpha}p_{j\alpha}\prod_{\alpha'\notin A_n}(1-p_{i\alpha'}p_{j\alpha'})\right].
\end{equation}
Since, in principle, every $p_{i\alpha}$ is different, the distribution described by Eq.~\ref{eq:PB} is a sequence of Bernoulli events, each with different probability and equal to the one expressed in Eq.~\ref{eq:p_v} and takes the name of Poisson-Binomial distribution~\cite{Hong2013}. 

Once we have the BiCM distribution for the number of co-occurrences between nodes $i$ and $j$, we can then calculate the statistical significance of the observed $V_{ij}^*$ via the p-value, i.e.

\begin{equation}\label{eq:p_vals}
\text{p-value}\big(V^*_{ij}\big)=\sum_{V_{ij}\ge V^*_{ij}}f_{PB}\big(V^*_{ij}\big).
\end{equation}

\ale{Iterating the calculation of Eq.~\ref{eq:p_vals} for every couple of nodes belonging to the $\top$ layer results in $\binom{N_\top}{2}$ p-values; to state the statistical significance of each of them, it is necessary to adopt a multiple hypothesis testing correction. In particular, the False Discovery Rate (FDR,~\cite{benjamini1995controlling}) is particularly effective since it permits to control the false positives rate.}\\

The procedure described in \ale{this} subsection was developed in Ref.~\cite{Saracco2017}. For the actual implementation, we used \href{https://bipartite-configuration-model.readthedocs.io/en/latest/}{\texttt{bicm}} python module, available on \href{https://pypi.org/project/bicm/}{\texttt{pypi}} and described as part of \href{https://nemtropy.readthedocs.io/en/master/index.html}{\texttt{NEMtropy}} package, in Ref.~\cite{vallarano2021fast}.
}

\aleY{\subsection*{Community detection on the validated projection network}
The choice of the community detection algorithm is not a trivial one, see for instance\cite{Fortunato2010,Peixoto2022}. In this case, we used Louvain as a descriptive method (using Peixoto's jargon\cite{Peixoto2022}), since we intended to \emph{describe} the mesoscale structure of the validated network of firms.\\ 
Since Louvain is known to be node-order dependent\cite{Fortunato2010}, we run the algorithm 1000 times after changing the order of the nodes, finally accepting the partition displaying the greatest value of the modularity. For completeness, we compared the Louvain partition with the ones coming from other models. In particular, we analysed the results from Infomap\cite{Rosvall2007} and WalkTrap\cite{Pons2005}, since they are built on a different rationale than the modularity used in Louvain~\cite{Fortunato2010}: for all these algorithms, we used the implementations present in the python module \href{https://python.igraph.org/en/stable/}{\texttt{python-igraph}}. Both algorithms return communities that are, on average, much smaller that the ones returned by Louvain. Moreover, the communities of InfoMap and WalkTrap are nearly completely embedded in the Louvain ones: on average, 82.52\% (99.14\%) of nodes in each InfoMap (WalkTrap) community belong to the same Louvain community. In a way, those methods are describing smaller structures than the ones captured by Louvain, that nevertheless, includes them. In our description, we intend to focus on particularly dense communities, since they represent groups of firms whose social communication is indeed similar:  a modularity-based algorithm fits with our aims.\\ 
In order to refine the description obtained, we rerun the Louvain algorithm inside each community: such an approach has the advantage of reducing the issues related to the resolution limit of modularity-based algorithms~\cite{Fortunato2010}.}

\begin{table}[ht!]
\centering
\begin{tabular}{|l|l|}
\hline
\hline
\textbf{NACE} & \textbf{descritpion}\\
\hline
\hline
A & Agriculture, forestry and fishing \\
\hline
B & Mining and quarrying\\
\hline
C & Manufacturing\\
\hline
D & Electricity, gas, steam and air conditioning supply\\
\hline
E & Water supply, sewerage, waste management and remediation activities\\
\hline
F & Construction\\
\hline
G & Wholesale and retail trade; repair of motor vehicles and motorcycles\\
\hline
H & Transportation and storage\\
\hline 
I & Accommodation and food service activities\\
\hline 
J & Information and communication\\
\hline
K & Financial and insurance activities\\
\hline
L & Real estate activities\\
\hline
M & Professional, scientific and technical activities\\
\hline
N & Administrative and support service activities\\
\hline
O &  Public administration and defence; compulsory social security\\
\hline
P & Education\\
\hline
Q & Human health and social work activities\\
\hline
R & Arts, entertainment and recreation\\
\hline
S & Other service activities\\
\hline
T & Activities of households as employers; undifferentiated goods- and services-producing
activities of households for own use\\
\hline
U & Activities of extra-territorial organisations and bodies\\
\hline
\end{tabular}
\caption{\textbf{NACE Rev.2, main division} The description of the categories was taken from Ref.~\cite{Eurostat2008}.\label{tab:nace}}
\end{table}

\ale{
\subsection*{Relating hashtags to SDGs}
To identify the SDG subjects that UK companies talk about, we used a threefold approach to have a good cover tailored to the available data set. As a list of SDGs-related keywords suitable for online social network searches does not exist, we had to create one. Considering the many attempts to map academic articles' contributions to SDGs, we first started from the list of the University of Auckland (\href{https://www.sdgmapping.auckland.ac.nz/}{available here}), which is used in business research \cite{Sinkovics2022} and consider the presence of words in our data set. This list mainly refers to keywords used in research papers in Elsevier's Scopus database, while in the present dataset, we are referring to Twitter's hashtags. In this sense, we gather multiple words in a single keyword, as it is customary for hashtags: for instance, ``Child Labor Laws" became ``childlaborlaws". Sometimes, the keywords were annotated under more than a single SDG: we disambiguated the multiple identifications manually, focusing on the main target of the various SDGs.
\\At this level, the identified SDG keywords represented less than $0.52\%$, i.e. quite a limited coverage. Since we were not aware if the limited coverage of SDG subjects was due to short attention to those arguments or not effective identification of SDG hashtags, we manually annotated the hashtags among the 300 most frequent ones related to an SDG. 
The two authors independently performed the identification and agreed on  $86.3\%$ of the annotations; when they did not agree on the hashtag categorisation, they discussed each hashtag and finally attributed an SDG when they reached an agreement.
Using this approach, we reached the $0.60\%$ of all hashtags used by accounts in the validated projection of Fig.~\ref{fig:valnetuser}.\\
To further enlarge the SDG covering, we used a network approach. Using the bipartite representation of accounts and hashtags already used to obtain the validated projection on the account layer, we projected the network on the layer of hashtags, using the technique described in the subsection above and introduced in Ref.~\cite{Saracco2017}. We remind the reader that in the validated projection, two nodes are present if they share a significant number of nearest neighbours in their bipartite representation. In this paper, two hashtags are connected in the validated projected network if they were both used by a significant number of different users. In this sense, a link in this network represents a non-trivial measure of similarity in how the various Twitter accounts use hashtags.\\ 
Some might argue that we are interested in hashtags appearing in the same messages. This point is debatable: an account interested in subjects related to, for example, SDG3 may use some of them related to different facets of SDG3 in different messages and focusing on hashtags used in the same messages will miss this information. Moreover, we avoid the risk of validating too many close hashtags since the procedure defined in~\cite{Saracco2017} is highly restrictive. For instance, in the hashtag-validated projected network, the link density is extremely low, i.e. 0.09\%.
Nevertheless, even in this case, we had to check the ``automatic" annotation manually: in fact, the (validated) link between two hashtags may be due to a different reason than the adherence to the aims of the SDG: for instance, the keyword \emph{\#worldengineeringday} is connected to only the hashtag \emph{\#inwed}, i.e. the acronym for the International Women in Engineering Day, but it is not necessarily related to Gender Inequality (SDG5), as its neighbour. In a way, the validated network represents a hint to spot possible SDG hashtags related to the already labelled ones. Moreover, it permits spotting SDG hashtags specific to the current data set. It is the case, for instance, of hashtags of the various campaigns of the National Health Systems (all of them have been classified in SDG3) that are not general or the ones related to the Covid19 vaccination.
We remark that in the validated network, the SDG hashtags represent a greater percentage (8\%), signalling that there is collective attention of company accounts on the various subjects.\\ 
We focused on all hashtags that were not assigned an SDG that have at least an SDG hashtag among their neighbours since we expect that the former hashtags are related to the SDG of their neighbours. To be more restrictive, we focus on hashtags whose neighbours that were assigned an SDG represented more than half of their degree. Then they were assigned the most frequent SDG in their neighbours. The association was later manually checked to manage the case of ties in the SDGs in the neighbours, resulting in 146 newly annotated hashtags. The annotated hashtags now represent 0.68\% of all hashtags in the data set. 
}






\section*{Author contributions statement}
AP: Conceptualization, Data curation, Writing – original draft, Writing – review \& editing
FS: Data curation, Formal analysis, Methodology, Visualization, Writing – original draft, Writing – review \& editing

\section*{Data Availability}
Twitter ID data can be downloaded from the following \href{https://drive.google.com/file/d/1XmS761hblQFnsg6ihynu2joMd5zdPjf7/view?usp=sharing}{link}. The list of hashtags associated to the various SDGs can be download from the following \href{https://drive.google.com/file/d/1ur--6Djo2YBCdzJx9-iYEoxSGW8GvD7C/view?usp=sharing}{link}.
The firms’ ID and financial data that support the findings of this study are available from AIDA (Bureau Van Dijk). Restrictions apply to the availability of these data, which were used under license for this study.

\section*{Competing interests}
The authors declare no competing interests.





\afterpage{%
    \clearpage
    \thispagestyle{empty}
    \begin{landscape}
    \small\addtolength{\tabcolsep}{-5pt}
        \centering 
        \begin{tabular}{||c|c||c|c||c|c||c|c||c|c||}
        \hline  \hline
        \multicolumn{2}{||c||}{\textbf{0}}&\multicolumn{2}{|c||}{\textbf{2}}&\multicolumn{2}{|c||}{\textbf{5}}&\multicolumn{2}{|c||}{\textbf{6}}&\multicolumn{2}{|c||}{\textbf{1}}\\
\hline
 hashtag               &   frequency & hashtag                &   frequency & hashtag        &   frequency & hashtag                &   frequency & hashtag                &   frequency \\
\hline \hline
 cloud                 &         151 & cop                    &         270 & covid          &          61 & covid                  &         147 & earthday               &          63 \\
 technology            &         140 & sustainability         &         259 & sustainability &          51 & cop                    &         101 & blackhistorymonth      &          53 \\
 digital               &         135 & netzero                &         248 & esg            &          50 & budget                 &          97 & pridemonth             &          53 \\
 data                  &         135 & climatechange          &         193 & cop            &          48 & webinar                &          94 & internationalwomensday &          50 \\
 digitaltransformation &         134 & sustainable            &         187 & china          &          40 & esg                    &          94 & covid                  &          50 \\
 cybersecurity         &         120 & earthday               &         182 & inflation      &          39 & internationalwomensday &          91 & iwd                    &          49 \\
 innovation            &         111 & internationalwomensday &         170 & supplychain    &          35 & brexit                 &          88 & sustainability         &          46 \\
 covid                 &         109 & innovation             &         148 & sustainable    &          34 & diversity              &          80 & pride                  &          42 \\
 tech                  &         106 & energy                 &         144 & innovation     &          34 & investment             &          79 & choosetochallenge      &          38 \\
 webinar               &         100 & worldenvironmentday    &         135 & climatechange  &          32 & podcast                &          78 & womenshistorymonth     &          36 \\
\hline
\hline
\end{tabular}
\normalsize
\captionof{table}{\textbf{Frequency of top 10 most frequent hashtags in the subcommunities with more than 50 nodes in the Cyan community.}}\label{tab:cyan_hashtags}

\small
\centering 
\begin{tabular}{||c|c||c|c||c|c||c|c||}
\hline \hline
        \multicolumn{2}{||c||}{\textbf{1}}&\multicolumn{2}{|c||}{\textbf{0}}&\multicolumn{2}{|c||}{\textbf{6}}&\multicolumn{2}{|c||}{\textbf{2}}\\
\hline
 hashtag                   & frequency   & hashtag                & frequency   & hashtag                   &   frequency & hashtag                   & frequency   \\
\hline\hline
 internationalwomensday    & 88          & christmas              & 97          & carehome                  &          42 & mentalhealthawarenessweek & 85          \\
 mentalhealthawarenessweek & 75          & valentinesday          & 88          & care                      &          41 & internationalwomensday    & 83          \\
 blackhistorymonth         & 58          & halloween              & 81          & christmas                 &          40 & covid                     & 63          \\
 pridemonth                & 52          & internationalwomensday & 80          & internationalwomensday    &          39 & mentalhealth              & 61          \\
 iwd                       & 44          & mondaymotivation       & 76          & covid                     &          37 & worldmentalhealthday      & 58          \\
 earthday                  & 43          & fridayfeeling          & 72          & socialcare                &          37 & volunteersweek            & 56          \\
 worldmentalhealthday      & 43          & mothersday             & 71          & dementia                  &          29 & christmas                 & 45          \\
 choosetochallenge         & 36          & win                    & 69          & halloween                 &          28 & blackhistorymonth         & 45          \\
 pride                     & 35          & easter                 & 67          & valentinesday             &          26 & wellbeing                 & 38          \\
 diwali                    & 33          & bankholiday            & 67          & mothersday                &          25 & worldbookday              & 38          \\
 cop                       & 33          &                        &             & mentalhealthawarenessweek &          25 &                           &             \\
                           &             &                        &             & internationalnursesday    &          25 &                           &             \\
\hline\hline
\end{tabular}
\normalsize
\captionof{table}{\textbf{ Frequency of top 10 hashtags  in the subcommunities with more than 50 nodes in Orange-red community, 1/2.}}\label{tab:orangered4_hashtags_1}

\small
\centering 
\begin{tabular}{||c|c||c|c||c|c||}
\hline
        \multicolumn{2}{||c||}{\textbf{5}}&\multicolumn{2}{|c||}{\textbf{4}}&\multicolumn{2}{|c||}{\textbf{3}}\\
\hline\hline
 hashtag                   & frequency   & hashtag                   & frequency   & hashtag                & frequency   \\
\hline
 covid                     & 43          & internationalwomensday    & 85          & christmas              & 34          \\
 learningdisability        & 41          & cop                       & 85          & volunteers             & 30          \\
 socialcare                & 35          & covid                     & 77          & volunteersweek         & 29          \\
 mentalhealthawarenessweek & 35          & blackhistorymonth         & 72          & charity                & 28          \\
 autism                    & 35          & mentalhealthawarenessweek & 71          & givingtuesday          & 25          \\
 internationalwomensday    & 32          & iwd                       & 66          & hospicecareweek        & 24          \\
 learningdisabilities      & 30          & earthday                  & 55          & internationalwomensday & 23          \\
 mentalhealth              & 28          & mentalhealth              & 50          & internationalnursesday & 22          \\
 learningdisabilityweek    & 25          & unimentalhealthday        & 50          & londonmarathon         & 22          \\
 autistic                  & 25          & diwali                    & 50          & fundraising            & 20          \\
                           &             &                           &             &                        &             \\
                           &             &                           &             &                        &             \\
\hline \hline
\end{tabular}
\normalsize
\captionof{table}{\textbf{ Frequency of top 10 hashtags  in the subcommunities with more than 50 nodes in Orange-red community, 2/2.}}\label{tab:orangered4_hashtags_2}

\small
\centering 
\begin{tabular}{||c|c||c|c||c|c||}
\hline \hline
        \multicolumn{2}{||c||}{\textbf{2}}&\multicolumn{2}{|c||}{\textbf{0}}&\multicolumn{2}{|c||}{\textbf{3}}\\
\hline
 hashtag                    & frequency   & hashtag                   &   frequency & hashtag                   & frequency   \\
\hline \hline
 apprenticeship             & 125         & worldbookday              &         147 & apprenticeship            & 92          \\
 mentalhealthawarenessweek  & 103         & antibullyingweek          &         111 & mentalhealthawarenessweek & 74          \\
 naw                        & 99          & mentalhealthawarenessweek &         109 & apprentice                & 72          \\
 internationalwomensday     & 97          & internationalwomensday    &          85 & internationalwomensday    & 67          \\
 apprentice                 & 83          & childrensmentalhealthweek &          85 & naw                       & 67          \\
 iwd                        & 64          & christmas                 &          80 & engineering               & 60          \\
 nationalapprenticeshipweek & 61          & remembranceday            &          77 & education                 & 53          \\
 mentalhealth               & 57          & backtoschool              &          71 & construction              & 52          \\
 choosetochallenge          & 55          & onekindword               &          70 & collegesweek              & 51          \\
 cop                        & 54          & saferinternetday          &          67 & careers                   & 51          \\
                            &             & science                   &          67 &                           &             \\
                            &             & wellbeing                 &          67 &                           &             \\
\hline \hline
\end{tabular}
\normalsize
\captionof{table}{\textbf{Frequency of top 10 hashtags  in the subcommunities with more than 50 nodes in Yellow community.}}\label{tab:yellow1_hashtags}
\small
\centering 
\begin{tabular}{||c|c||c|c||c|c||c|c||}
\hline \hline
\multicolumn{2}{||c||}{\textbf{3}}&\multicolumn{2}{|c||}{\textbf{0}}&\multicolumn{2}{|c||}{\textbf{2}}&\multicolumn{2}{|c||}{\textbf{1}}\\
\hline
 hashtag                   & frequency   & hashtag                   & frequency   & hashtag                   &   frequency & hashtag                   &   frequency \\
\hline \hline
 covid                     & 128         & mentalhealthawarenessweek & 47          & covid                     &         106 & covid                     &          55 \\
 nhs                       & 123         & covid                     & 43          & covidvaccine              &          81 & mentalhealthawarenessweek &          40 \\
 covidvaccine              & 111         & nhs                       & 36          & mentalhealthawarenessweek &          79 & internationalwomensday    &          38 \\
 internationalnursesday    & 103         & mentalhealth              & 35          & volunteersweek            &          67 & internationalnursesday    &          38 \\
 volunteersweek            & 89          & internationalwomensday    & 34          & internationalwomensday    &          62 & nhs                       &          36 \\
 ahpsday                   & 86          & covidvaccine              & 34          & euro2020                  &          58 & covidvaccine              &          33 \\
 blackhistorymonth         & 84          & worldmentalhealthday      & 33          & worldmentalhealthday      &          57 & worldmentalhealthday      &          32 \\
 internationalwomensday    & 83          & volunteersweek            & 32          & nhs                       &          55 & timetotalk                &          26 \\
 mentalhealthawarenessweek & 81          & blackhistorymonth         & 32          & grabajab                  &          54 & vaccine                   &          26 \\
 nhsbirthday               & 80          & wellbeing                 & 32          & everymindmatters          &          49 & mentalhealth              &          25 \\
                           &             &                           &             & mentalhealth              &          49 & grabajab                  &          25 \\
\hline \hline
\end{tabular}
\normalsize
\captionof{table}{\textbf{Frequency of top 10 hashtags  in the subcommunities with more than 50 nodes in Orchid community.}}\label{tab:orchid_hashtags}

    \end{landscape}
    \clearpage
}

\end{document}